\documentclass{article}
\usepackage{graphicx}
\usepackage{cite}
\usepackage{amssymb,amsmath}

\DeclareGraphicsExtensions{.eps,.jpg,.pdf,.ps} 

\begin{document}

\bibliographystyle{plain}

\title{Rough surface backscatter and statistics 
via extended parabolic integral equation}
\author{Mark Spivack and Orsola Rath Spivack}

\font\eightrm=cmr8

\date{\today}

\maketitle

\vskip 1 true cm
{\eightrm
\begin{center}
Department of Applied Mathematics and Theoretical Physics,
The University of Cambridge CB3 0WA, UK

\end{center}
}

\vskip 1.5 true cm
\begin{abstract}
This paper extends the parabolic integral equation method, which is
very effective for forward scattering from rough surfaces, to include
backscatter. 
This is done by applying left-right splitting to a modified
two-way governing integral operator, to express the solution as a series of
Volterra operators; this series describes
successively higher-order surface interactions between
forward and backward going components, and allows highly 
efficient numerical evaluation.  This and equivalent methods such as
ordered multiple interactions have been developed for
the full Helmholtz integral equations, but not previously applied to the
parabolic Green's function. 
In addition, the form of this Green's function allows the mean field
and autocorrelation to be found analytically 
to second order in surface height.  These may be regarded as backscatter
corrections to the standard parabolic integral equation
method.

\end{abstract}

\def\EQN{\eqno}
\def\beq{\begin{equation}}
\def\eeq{\end{equation}}
\def\beqn{\begin{eqnarray}}
\def\eeqn{\end{eqnarray}}
\def\bef{\begin{figure}}
\def\eef{\end{figure}}

\vfil\break

\section {Introduction}\label{intro}

Wave scattering from irregular
surfaces continues to present formidable theoretical and computational
challenges \cite{warnick,zhang,1,3,vor1,saillard,4},
especially with regard to analytical treatment of
statistics, and numerical solution for wave
incidence at low grazing angles \cite{chan6,7,8,9,sum,qi}, where the
insonified/illuminated region may become very large.
Computationally, the cost of the necessary matrix inversion scales
badly with wavelength and domain size and can rapidly become
prohibitive; 
this is compounded by the large number of Green's
function evaluations, whose overall cost is therefore sensitive to the
form which this function takes.

Under the assumption of purely forward-scattering, 
a successful approach has been the
parabolic integral equation method (PIE)\cite{10,11,12a}. This makes use of  
a `one-way' parabolic equation (PE) Green's function, 
leading to the replacement of the Helmholtz integral equations by
their small-angle analogue.
For 2D problems this Green's function takes a particularly tractable form;
this, together with the Volterra (one-sided) form of the governing
integral operator, affords the key advantage of high
numerical efficiency, and in the perturbation regime allows
derivation of analytical 
results \cite{12,13,14,16}.
Nevertheless, the method yields
no information about the field scattered back towards the source.

On the other hand, where backscatter is required, 
operator series solution methods such as left-right splitting and method of
ordered multiple interactions
\cite{kapp1,spivack3,spivack1,tran,adams,spivack4,pino}
have proved highly versatile, in both 2 and 3 dimensions.
These use the full free-space Green's function and
proceed by expanding the surface fields about the dominant
`forward-going' component, and thereby circumvent the difficulties of
tackling the full Helmholtz equations.

In this paper we combine these approaches, extending the standard PIE
description to a `two-way' method, thus allowing for
both left- and right-travelling waves.
This is obtained in the obvious way by replacing the parabolic
equation Green's function by a form symmetrical in range.
The integral operator can be
split into left- and right-going parts; under the
assumption that forward scattering dominates,
the solution can then be written
as a series and truncated. Every term of this series is a product of
Volterra operators and is therefore treated as efficiently as the standard PIE
method, which corresponds approximately\footnote
{Note however that in contrast to standard PIE the first term includes
`direct backscatter' without additional effort.}
 to truncation at the first term.

In the second part of the paper we impose the additional restriction
to the perturbation regime of small surface height $\sigma$, within which
analytical expressions for the mean field and autocorrelation function
are obtained. This extends 
the corresponding results\cite{12,13} derived under the PIE method.
The approach there was first 
to obtain the scattered field to second order in $\sigma$ at the mean
surface plane, and find the far-field under the assumption that
propagation outwards from the surface is governed by the full Helmholtz
equation. In the standard PE case,
this modification allows a small amount of backscatter, but precludes any
backscatter enhancement which can be thought of as due to coherent addition of
reversible paths\cite{18,19,20},
because interactions at the surface are assumed to be take
place in the forward  direction only.
The formulation presented here allows one to remove
this restriction, and separate the forward and backward going
interactions to various orders, although this aspect is not explored
in detail here. 
In particular this method produces a correction
term, whose statistics can be obtained in the perturbation regime.

The paper is organised as follows: The standard parabolic integral
equation method and 
preliminary results are given in section \ref{I}.  In section \ref{II} the full
two-way parabolic integral equation method is set out, and the iterative
solution explained. 
Analytical results for the statistics under the extended
method are derived in section \ref{III}.

\section{Parabolic integral equation method and
  preliminaries}\label{I} 

\bigbreak

\begin{figure}
\hskip 1 true cm
\includegraphics[height=6cm]{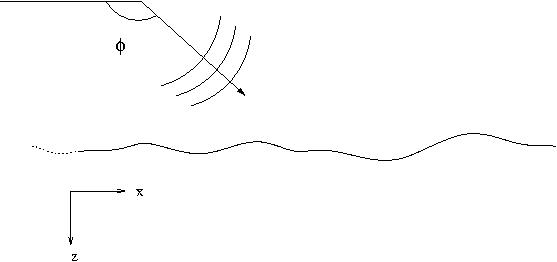} 
\caption{\label{Fig.1} Schematic view of scattering geometry}
\end{figure}

We consider the problem of a scalar time-harmonic wave field $p$ scattered
from a one-dimensional rough surface $h(x)$ with a pressure release boundary
condition. (Equivalently, $p$ is an electromagnetic $s$ or $TE$
polarised wave and $h$ is a perfectly conducting corrugated surface whose
generator is in the plane of incidence.)
The wavefield has wavenumber $k$ and is
governed by the wave equation $(\nabla^2+k^2)p=0$. The coordinate axes are $x$
and $z$ where $x$ is the horizontal and $z$ is the vertical, directed out of
the medium (see Fig. \ref{Fig.1}). Angles of incidence and scatter are assumed to be
small with respect to the positive $x$-direction. It will be assumed that the
surface is statistically stationary to second order, i.e. its mean and
autocorrelation function are translationally invariant. We may choose
coordinates so that
$h(x)$ has mean zero. The autocorrelation function
$<h(x)h(x+\xi)>$ is denoted by $\rho(\xi)$, and we assume that
$\rho(\xi)\rightarrow 0$ at large separations $\xi$.
(The angled brackets here denote the ensemble average.)
Then $\sigma^2 \equiv \rho(0)$ is the variance of surface height, so that the
surface roughness is of order $O(\sigma)$. 

Since the field components
propagate predominantly around the $x$-direction, we can define
a slowly-varying part $\psi$ by
\beq \psi (x,z) ~=~ p(x,z) ~\exp (-ikx)  .\eeq 
Slowly varying
incident and scattered components $\psi_i$ and $\psi_s$ are defined
similarly, so that $\psi= \psi_i + \psi_s$.
It may be assumed that $\psi_i (x,h(x)) = 0$
for $x \leq 0$, so that the area of
surface insonification is restricted, as it would be for example in the case
of a directed Gaussian beam.
The governing equations for the standard parabolic equation method
\cite{10,11}  
are then

\beq  \psi_i ({\bf r}_s) = - \int_0^x  ~G_p ({\bf r}_s; {\bf r'}) 
\frac{  \partial \psi({\bf r'}) } {{\partial z} }~dx' \label{1} \eeq 
where both ${\bf r}_s=(x,h(x))$, ${\bf r'}=(x',h(x'))$ lie on the surface; and

\beq  \psi_s({\bf r}) = \int_0^x ~G_p ({\bf r}; {\bf r'}) 
\frac{ \partial \psi({\bf r'}) } {{\partial z}  }~dx' \label{2} \eeq 
where ${\bf r'}$ is again on the surface and ${\bf r}$ is an arbitrary
point in the medium.  Here $G_p$ is the parabolic form of the
Green's function in two dimensions given by
\[ G_p(x,z; x', z')  \left\{ 
\begin{array}{lr} 
 = & \alpha~{\sqrt{ 1 \over  x - x'}}~
\exp\Bigl[{ik (z - z')^2 \over 2(x - x')}\Bigr] \hfill~~~~~~{\rm for~}
x'~<~x  \\
= & 0  \hfill~~~~~~~~~~~~~~{\rm otherwise}  
\label{3}
\end{array} 
\right.
\]
where $\alpha = {1\over 2}~{\sqrt{i/ 2\pi k}}$.  This asymmetrical form
gives rise to the
finite upper limit of integration in (\ref{1}) and (\ref{2}).
It is derived under the assumption of forward-scattering,
and that the field obeys the parabolic wave equation,
\beq \psi_x ~+~ 2 i k \psi_{zz} ~=~ 0 \label{4} \eeq 
which holds provided the angles of incidence and scattering are fairly small
with respect to the $x$-direction.
($G_p$ can also be obtained directly from the full free space Green's function
under the small-angle approximation.)
Equation (\ref{1}) must be inverted to give the induced source
$\partial\psi/\partial z$ at the surface, which is then
substituted in (\ref{2}) to determine the field elsewhere.

Now, equations (\ref{1}) and (\ref{2}) do not apply to plane wave scattering
at small or negative $x$
because of the truncated lower limit of integration, equivalent to the
restricted surface insonification.  Nevertheless, we can formally apply the
integral equation to a plane wave,
to obtain a solution which will be
physically meaningful and asymptotically accurate at large values of $x$.
This procedure has been used\cite{12,13} to derive the field statistics;
where necessary we will
assume that $x$ is sufficiently large for this to hold.

Consider an incident plane wave $p=\exp(ik[x\sin\theta + z\cos\theta])$, where
$\theta$ is the angle with respect to the vertical. The grazing angle is
then denoted $\mu=\pi/2-\theta$ (see Fig. \ref{Fig.1}).
This plane wave has slowly-varying component $\psi^\theta=\exp(ik[Sx +
z\cos\theta])$, where
\beq  S ~=~ \sin\theta - 1 , \label{5} \eeq 
which we refer to as the reduced plane wave.

\vfil\break
\section{Two-way parabolic integral equation method}\label{II}

In this section the two-way version of the PIE method will be described, and
the iterative solution will be given. This provides an efficient means of
calculating the back-scattered component at small angles of scatter.

\subsection{The modified governing equations}

\def\psinc{\psi_i}
\def\psiz{{\partial\psi\over \partial z}}
\def\psizz{{\tilde \psi}}
\def\rbar{{\bf r}}
The governing equations (\ref{1}), (\ref{2}) must first be modified
to take into account scattering from the right.  To do this, we simply
replace
$G_p$ by its symmetrical analogue $G$.
This form arises if we apply the small
angle approximation described in section \ref{I}
to the full free space Green's function without requiring
$G(x,z;x',z')$ to vanish when $x'\geq x$. We thus obtain

\beq
G(x,z; x', z')  ~ \left\{ 
\begin{array}{lr} 
~&=~
\alpha~{\sqrt{ 1 \over  x - x'}}~
\exp\Bigl[{ik (z - z')^2 \over 2(x - x')}\Bigr] \hfill~,
~~~~~~~~~~~~~~~~~~~~~~x'~<~x  \cr
~&=~
\alpha~{\sqrt{ 1 \over  x' - x}}~
\exp\Bigl[{ik (z - z')^2 \over 2(x' - x)}\Bigr]~\exp\left[2ik(x'-x)\right] ~
~~~~~~
x'~\geq~x 
\label{6}
\end{array}
\right.
\eeq
The factor $\exp[-2ik(x'-x)]$ arises for $x' \geq x$
because we are solving for the reduced
wave $\psi$.

Applying this Green's function
to the reduced wave $\psi$ we obtain
\beq  \psi_s (x,z) =  \int_0^\infty G(\rbar,\rbar')
{\partial\psi (\rbar')\over \partial z} dx'
\label{7} \eeq 
where $\rbar=(x,z)$, $\rbar'=(x',h(x'))$.
This is the analogue of equation
(2), effectively containing a back-scatter correction.
Taking the limit of (7) as $z \rightarrow h(x)$ yields an
integral equation relating the incident field to the scattered field at the
surface:
\beq  \psinc (x,h(x)) = -\int_0^\infty G(\rbar_s,\rbar')
{\partial\psi (\rbar')\over \partial z} dx'
\label{8} \eeq 
where now $\rbar_s=(x,h(x))$, $\rbar'=(x',h(x'))$ both lie on the surface.
(Note that the addition of a correction to the parabolic equation is
closely related to a method proposed by Thorsos\cite{10}.)
Equations (7), (8) can be written in operator notation:
\beq  \psi_s (x,z) = -(L+R){\partial\psi\over \partial z} \label{9} \eeq 
\beq  \psinc (x,h(x)) = (L+R){\partial\psi\over \partial z} \label{10} \eeq 
where $L$, $R$ are defined by
\beq Lf(x,z)=\int_0^x G(\rbar,\rbar') f(x') dx',
~~~~~ Rf(x,z)=\int_x^\infty G(\rbar,\rbar') f(x') dx'  \nonumber\eeq 
and $\rbar=(x,z)$, $\rbar'=(x',h(x'))$. These integral operators
and their inverses are Volterra, or `one-sided' in an obvious sense.

\subsection{Solution of the modified equations}

The main computational task in any such boundary integral method is the
inversion of the integral equation (\ref{10}).
One of the principal advantages of the
standard forward-going PIE method (equations (\ref{1})-(\ref{2}))
is that its one-way form allows Gaussian elimination to be
used, so that inversion is highly efficient.
In the above two-way formulation this
advantage is initially lost, since direct inversion of $L+R$ in eq. (\ref{10})
offers no benefit compared with
solving the full Helmholtz equations.  However, the 
computational advantage can be regained
by forming an iterative series solution, in which each term is a
product of Volterra integral operators.

Integral equation (\ref{10}) has formal solution
\beq  {\partial\psi\over \partial z}  = (L+R)^{-1}\psinc   \label{11} \eeq 
which can be expanded in a series
\beq  \psiz = \left[L^{-1} - L^{-1} R L^{-1} + \left(L^{-1} R\right)^2 L^{-1} -
...\right] \psinc \label{12} \eeq 

Under the assumption that $R$ is small in the following sense
the series (\ref{12}) is convergent, as is already 
required implicitly for the standard PIE solution;
the series can then be
truncated after finitely many terms.
By `small' we mean that $R\phi/||\phi||$ is small for all terms $\phi$
in the series.  It can be shown that this assumption
is indeed justified at low grazing angles
for surfaces whose slopes are not too large,
since the kernel of $R$ oscillates rapidly especially at small wavelengths.
It is nevertheless difficult to  give this a precise range of validity, and we
will not attempt to do so here.

Solution for the field can therefore be obtained by truncating the 
series (\ref{12}) and substituting into the integral (\ref{7}).
The {\it first term} $L^{-1}\psinc$ in series (\ref{12}) 
corresponds to the solution for $\partial\psi/\partial z$
under the standard PIE method (e.g. \cite{11}).
Denote this first approximation by $\psizz$, i.e.
\beq \psiz \cong \psizz = L^{-1} \psinc . \label{13} \eeq 
Note however that the integral (\ref{7}) allows 
for outgoing components scattered to the left, unlike its PIE analogue
(\ref{2}),
so even this lowest order truncation gives backscatter.
This can be considered the {\it direct backscatter} component.

Truncation of (\ref{12}) at the {\it second term} gives:
\beq  {\partial\psi\over \partial z} \cong  \psizz + C
\label{14} \eeq 
where $C$ is a correction term,
\beq C= L^{-1}RL^{-1}\psi_i . \label{15} \eeq 
The above expression will be used in section \ref{III} to obtain
some statistical measure of the backscattered component in the perturbation
regime of small surface height.
We remark that this is the lowest-order truncation consistent with
reversible ray paths.

\subsection{Numerical evaluation}
The general term of (\ref{12}) is a product of the operators $L^{-1}$ and $R$.
Evaluation of the integral $R$ is straightforward.
For computational purposes we assume that the incident wave insonifies
only a finite region of the rough surface; the source may for example be
a Gaussian beam. A finite upper limit of integration $x_{max}$, say, may
then be assumed.

Numerical inversion of $L$ is also highly efficient since
discretization of $L$ gives rise to a lower-triangular matrix. This
has been described
elsewhere (e.g. \cite{11}) and will only be summarized here.

Consider the equation $L\partial\psi/\partial z=\psi_i$ obtained by 
truncating (\ref{12}) at the first term.
This equation is discretized with respect to range $x$ using, say,
$N$ equally spaced points $x_j$. This then yields a matrix equation
$A\partial\psi/\partial z=\psi_i$ in which the matrix $A$ is
lower-triangular. Numerical inversion of this expression is carried out by
Gaussian elimination, requiring $O(N^2)$ operations, which
compares with $O(N^3)$ operations required to treat the full Helmholtz integral
equation.

The solution is thereby obtained for the first term, $\psizz$.
Typically only one further term, $L^{-1}R\psizz$, will be required.
The simplest way to obtain this is to discretize the integral $R$,
evaluate $R\psizz$ numerically, and then to solve
\beq  L^{-1}R\psizz = {\partial \psi\over \partial z} - \psizz 
\nonumber \eeq 
by Gaussian elimination as before.
The evaluation of the integral $R$ also requires $O(N^2)$ operations.
Subsequent terms in the series may be obtained similarly.

The computation can be simplified further in the perturbation regime
of small scaled surface height $k\sigma$,
if the operators $L$ and $R$ are replaced by the flat surface forms
in the calculation of the correction term $C$.  This is described in the
section \ref{III}.

\section{Perturbation solution and statistics of
  backscatter}\label{III} 

\subsection{Perturbation solution}
The mean field and higher moments based
on the standard parabolic equation approximation were obtained
elsewhere\cite{12,13} to second order in surface height in the case of
pure forward scattering. In this section the statistics of the
backscatter correction (eq. (\ref{14})) due to the two-way PIE method
will be derived. 

Suppose that a reduced  plane wave
$\psi_i^\theta=\exp(ik[xS+z\cos\theta])$ is incident on the rough surface at 
an angle $\theta$ measured from the normal.
We first summarize the perturbational calculation used to obtain the 
scattered field statistics previously.
Suppose that a plane
$z=z_1$, say, can be chosen `close' to every point on the surface.
The scattered
field is obtained to second order in surface height along this plane, for a
given incident plane wave, and the statistics are found from this.  Statistical
results obtained in this way do not depend on the choice of $z_1$ so for
convenience we may set $z_1=0$.
An expression is thus found for the scattered field
\beq  \psi_s(x,0) ~=~ -\psi_i^\theta (x,h) - h \left[\psiz -
{\partial \psi_i \over\partial z}\right] - {1\over 2} h^2
{\partial^2 \psi_i(x,0) \over\partial z^2} ~+~ O(\sigma^3) . \label{16} \eeq 
The only term here which is 
not known {\it a priori} is $\partial\psi/\partial z$.
The standard PIE solution $\psizz$ for $\partial\psi/\partial z$
is given\cite{12, 13} to second order in $\sigma$ by:

\beq \psiz \cong \psizz = -
{1\over \pi}~{d ~\over dx}\int_0^x ~{\psi_i^\theta (x',h(x'))\over
\alpha\sqrt{x-x'}}~dx' . \label{17} \eeq 
This arises from (\ref{13}) by substitution of the flat surface form
of $L$ (see (\ref{20}) below).
Denote by ${\tilde \psi_s}$
the approximation to $\psi_s$
obtained by substituting (\ref{17}) in (\ref{16}), so that
\beq
\begin{aligned}  \tilde\psi_s(x,0) ~\equiv~
~ - ~\psi_i^\theta (x, h) &+ h~ \biggl[
{1\over \pi}~{d ~\over dx}\int_0^x ~{\psi_i^\theta (x',h(x'))\over
\alpha\sqrt{x-x'}}~dx'
~+~ {\partial\psi_i^\theta (x, h)\over \partial z} \biggr] \\
~&-~ {h^2\over 2}{\partial^2\psi_i^\theta (x, 0)\over \partial z^2}
.  \label{18} \end{aligned} 
\eeq

We wish to calculate the
backscatter correction to this expression
due to the replacement of $\partial\psi/\partial z$
in (\ref{16}) by the corrected two-way PE solution
$(\psizz + C)$ (equations
(\ref{14}), (\ref{15})). We therefore repeat the above derivation  replacing
(\ref{13}) by (\ref{14}), to obtain
\beq  \psi_s (x,0) = {\tilde \psi_s}(x,0) + h(x) C(x) . \label{19} \eeq 
Since the correction term $C$ appears here with a factor $h$,
it is necessary to evaluate it
only to order $O(\sigma)$.

Expanding $L$ and $R$ (eqs. (\ref{9})-(\ref{10})) in surface height
$h(x)$, it is seen that $L=L_0+O(\sigma^2)$,
 $R=R_0+O(\sigma^2)$, where $L_0$, $R_0$
denote the deterministic (i.e. flat surface) forms of the operators $L$
and $R$ respectively:
\beq  L_0 = \alpha \int_0^x {1\over\sqrt{x-x'}} ~ dx' ,
~~~~~ R_0 = \alpha \int_x^\infty {1\over\sqrt{x'-x}} ~ dx' \label{20} \eeq 
In evaluating $C$ (eq. (\ref{15})) to order $O(\sigma)$ we may thus ignore
fluctuating parts of the operators, and replace $L$, $R$ by $L_0$, $R_0$
respectively.
We can therefore write
\beq  C = L_0^{-1} R_0 \psizz + O(\sigma^2)  .  \label{21} \eeq 
An expression of the form
$f=L_0^{-1}g$ is Abel's integral equation, which has the
well-known solution\cite{17a}
\beq  g(x) = {1\over \alpha\pi}{d\over dx} \int_0^x {1\over\sqrt{x-y}}
f(y) dy . \nonumber \eeq  
Now to first order in $h$, $\psizz$ in (21) is given\cite{12} by
\beq  \psizz (r) \cong -\pi\left[D_\theta (r) + {dI(r)\over dr}\right] .
\label{22} \eeq 
where, for large $r$,
$D$ takes the form (see eq. (\ref{15}) of \cite{12})
\beq  D_\theta (r) \sim -2ik \pi \sqrt{2-2\sin\theta} e^{ikSr} \label{23} \eeq 
and $I$ is an integral
\beq  I(r) ~=~ ~\int_0^r
~ikh(r')\cos\theta{e^{ikSr'}\over \alpha\sqrt{r-r'}}~dr' . \label{24} \eeq 
Therefore $D$ and $dI/dr$ are $O(1)$ and $O(h)$ respectively, so that
in eq. (\ref{21}) $C$ becomes
\beq  C(x) = {1\over\alpha^2\pi} {d\over dx}\left[\int_0^x
{1\over\sqrt{x-y}}
\int_y^\infty {\exp(ikr)\over\sqrt{y-r}} \psizz (r) ~dr ~dy
\right] . \label{25} \eeq 
To second order in surface height the scattered field $\psi_s(x,0)$ at the
mean surface is therefore described by eq. (\ref{19}), with $C$ given
by (\ref{25}). 

\subsection{Mean field}
\def\CC{{\cal E}}
The effect of the correction term $C$ on the scattered field statistics 
can now be examined. We
first find the mean field $<\psi_s(x,z)>$.
It is sufficient to obtain this quantity on the mean surface plane $z=0$, using
equation (\ref{19}), i.e.
\beq  <\psi_s(x,0)> ~=~ <\tilde\psi_s(x)> + <h(x)C(x)>  . \nonumber \eeq 
The solution for $<\tilde\psi_s>$ has been obtained
previously\cite{12}, and we can restrict attention to finding
the correction $<hC>$ to this.
Denote the correlation $<h(X)C(x)>$ by $\CC$ for any $X$, $x$, i.e.
\beq \CC(X,x)=<h(X)C(x)> . \nonumber \eeq 

Consider first the function $<h\psizz>$.
Since $<hD_\theta>$ vanishes, eq. (\ref{22}) gives
\beq <h\psizz> = -\pi <h{\partial I\over\partial x}>  . \label{26} \eeq 
Now from eq. (\ref{25})
\beq  \CC (X,x) =  \left<
{h(X)\over\alpha^2\pi}~ {d\over dx}\int_0^x {1\over\sqrt{x-y}}
\int_y^\infty {\exp(ikr)\over\sqrt{y-r}} \psizz (r) ~dr~
dy  \right>  . \nonumber \eeq 
The term $h(X)$ can be taken under the integral signs as part of
the operand of $d/dx$.
The order of integration and averaging can then be reversed so that, by
(26),
\beq  \CC (X,x) =
-{1\over\alpha^2} \left[{d\over dx}\int_0^x {1\over\sqrt{x-y}}
\int_y^\infty {e^{ikr}\over\sqrt{y-r}}
\left< h(X){\partial I(r)\over\partial r}  \right>
~ dr ~ dy \right]  . \label{27} \eeq 
Consider the term $<h(X) dI/dr>$ in the inner integrand.
By (\ref{24}),
\beq
 \begin{aligned} \left< h(X) {\partial I(r)\over\partial r} \right> =&
 \left< h(X) {d\over dr}
 ~\int_0^r
 ~ikh(r')\cos\theta{e^{ikSr'}\over \alpha\sqrt{r-r'}}~dr' \right> \\
 ~=& ik\cos\theta {d\over dr} \left[
 ~\int_0^r
 ~{e^{ikSr'}\over \alpha\sqrt{r-r'}}\rho(X-r')~dr' \right] 
 \label{28} 
 \end{aligned} 
\eeq
This may be substituted into (\ref{27}) to
give an analytical expression for the correlation $<h(X)C(x)>$. We can
simplify this expression by evaluating the
derivatives explicitly.
The term $\rho(X-r')$ is independent of $r$, so writing
\beq  {e^{ikSr'}\over \alpha\sqrt{r-r'}}=f(r,r') \label{29} \eeq 
the expression (\ref{28}) becomes
\beq  \left< h(X) {\partial I(r)\over \partial r}\right> =
ik\cos\theta {d\over dr} \left[
~\int_0^r
~f(r,r')\rho(X-r')~dr' \right]  \nonumber \eeq 
\begin{align}  =&
ik\cos\theta ~\lim_{\epsilon\rightarrow 0} {1\over\epsilon}\left[
\int_0^{r+\epsilon}
f(r+\epsilon,r')\rho(X-r')dr' -
\int_0^r f(r,r')\rho(X-r') dr' \right] \label{30} \\
=&
ik\cos\theta ~\lim_{\epsilon\rightarrow 0} {1\over\epsilon}
 \left[K_1 + K_2 - K_3 \right]   \nonumber
\end{align} 
where

\begin{align}  
K_1 &= \int_0^\epsilon f(r+\epsilon,r')\rho(X-r')dr' \nonumber \\
K_2 &= \int_\epsilon^{r+\epsilon} f(r+\epsilon,r')\rho(X-r')dr' \\
K_3 &= \int_0^r f(r,r')\rho(X-r') dr' \nonumber  \label{31} 
\end{align} 
Consider these three integrals in detail.
The first gives
\beq  {1\over\epsilon} K_1=
{1\over\epsilon}
\int_0^{\epsilon} f(r+\epsilon,r')\rho(X-r')dr' \cong {1\over\epsilon}
\rho(X)\int_0^{\epsilon}
{1\over\sqrt{r+\epsilon-r'}} dr' \nonumber\eeq 
\vskip -0.8cm
\begin{align}  = &
{2\over\alpha\epsilon}
\rho(X)\left[\sqrt{r+\epsilon}-\sqrt{r}\right]  
\label{32} \\
 \cong & {\rho(X) \over \alpha\sqrt{r}} \nonumber
\end{align} 
using a Taylor expansion in $\epsilon$.
Changing variables, $K_2$ in can be written
\beq  \int_\epsilon^{r+\epsilon} f(r+\epsilon,r')\rho(X-r')dr'
= \int_0^r f(r+\epsilon,r''+\epsilon) \rho(X-r''-\epsilon) dr'' . 
\label{33} 
\eeq
Now
\beq f(r+\epsilon,r''+\epsilon)={e^{ikS(r''+\epsilon)}\over \alpha
\sqrt{r-r''}}
 = e^{ikS\epsilon}f(r,r'' \nonumber\eeq 
so from (\ref{33})
\beq   \int_\epsilon^{r+\epsilon} f(r+\epsilon,r')\rho(X-r')dr'
= \int_0^r e^{ikS\epsilon} f(r,r')\rho(X-r'-\epsilon) dr' . \label{34} \eeq 
Thus the difference $K_2-K_3$ in (\ref{30}) becomes
\beq  
\begin{aligned} &\int_0^r f(r,r') \left[ e^{ikS\epsilon}
\rho(X-r'-\epsilon)-\rho(X-r')\right] dr'\\ 
 \cong &
\int_0^r f(r,r') \epsilon \left[ ikS
\rho(X-r')-{d\rho(X-r')\over dX}
\right] dr'
\label{35} 
\end{aligned}
\eeq 
where $\rho$, which may be assumed to be differentiable, has been expanded
to leading order in $\epsilon$.
Substituting (\ref{32}) and (\ref{35}) in (\ref{28}),
we obtain
\beq \left< h(X){\partial I(r)\over\partial r}  \right>
= {ik\over\alpha}\cos\theta \left\{  {\rho(X)\over\sqrt{r}} +
\int_0^r {e^{ikSr'}\over\sqrt{r-r'}} \left[
ikS\rho(X-r')-{d\rho(X-r')\over dX} \right] dr' \right\}
. \label{36} \eeq 
This removes the derivative with respect to $x$ in (\ref{27}), and indeed for
several important autocorrelation functions eq.~(\ref{36}) can be written
in closed form.  The term $\rho(X)/\sqrt{r}$
is an artifact of the finite lower bound of integration and
can be dropped, as we can assume the range variable $X$
to be large.
Equation (\ref{27}) therefore becomes
\beq  \CC (X,x) \equiv \left< h(x) C(x) \right> =
-{ik\over\alpha^3}\cos\theta \times \nonumber \eeq 
\beq  \left[{d\over dx}\int_0^x
{1\over\sqrt{x-y}}
\int_y^\infty {e^{ikr}\over\sqrt{y-r}}
\int_0^r {e^{ikSr'}\over\sqrt{r-r'}}
R(X,r') dr'
~ dr ~ dy \right]   \label{37} \eeq 
where
\beq  R(X,r')=ikS\rho(X-r')-{d\rho(X-r')\over dX} . \label{38} \eeq

The derivative with respect to $x$ in (\ref{37}) can
be evaluated similarly,
and after further manipulation (see Appendix)
the required expression can be written, setting $X=x$,
\beq  \left< h(x) C(x) \right> = -{ik\over\alpha^3} \cos\theta~
\times 
\nonumber
\eeq 
\beq
\begin{aligned}   &
\left[ {2\over\sqrt{x}}\int_y^\infty {e^{ikr}\over\sqrt{y-r}} \int_0^r
{e^{ikSr'}\over\sqrt{r-r'}} R(X,r') dr' dr
\right.   \\
& - \left. \int_0^x {1\over\sqrt{x-y}}
\int_y^\infty {e^{ikr}\over\sqrt{y-r}} \int_0^r
{e^{ikSr'}\over\sqrt{r-r'}}
{\cal F}(X,r') ~ dr' ~ dr ~dx \right]_{X=x}
 \label{39} \end{aligned} 
\eeq
where
\beq  {\cal F} = \left\{(1+ik\sin\theta)R(X,r') + {dR\over dr'}\right\} .
\label{40} \eeq 

\subsection{Autocorrelation and angular spectrum}
The main quantity of interest is the angular spectrum of intensity,
which may be defined as the Fourier transform of the autocorrelation function
(i.e. the second moment) of the scattered field.
This remains essentially unchanged with distance from the surface,
so that we may again concentrate on obtaining the form on the mean surface
plane, $z=0$.

Denote the second moment
\beq  m_2(x,y)= \left< \psi_s (x,0) \psi_s^*(y,0) \right> \nonumber \eeq 
where $^*$ indicates the complex conjugate,
and denote its approximation using the standard parabolic equation method by
\def\mtil{{\tilde m_2}}
\def\psizs{{\tilde \psi_s}}
\beq \mtil(x,y) \equiv \left< \tilde\psi_s (x,0) \tilde\psi_s^*(y,0)
\right> .\nonumber \eeq  
The perturbational solution of $\mtil$ was obtained in \cite{13}.
It is relatively straightforward to express $m_2$, to second order
in surface height under the present two-way PIE method,
as the sum of $\mtil$ and correction terms.
These additional terms, which are expected to be small,
represent the `indirect' contribution to the backscatter. 

From (\ref{19}) we have
\beq \psi_s (x) \psi_s^*(y)  = \psizs (x) \psizs^*(y) + \psizs(x) h(y) C^*(y)
+ \psizs^*(y) h(x) C(x) + h(x)h(y)C(x)C^*(y) . \label{41} \eeq 
We can write $\psizs$ and $C$ to zero and first order in surface height,
\beq  
\psizs=\psi_0 + \psi_1  + O(\sigma^2) 
\nonumber
\eeq 

where \cite{12}
\beq
\begin{aligned}  
\psi_0 (x) &= - e^{ikSx}   \\
    \psi_1 (x) &= - 2ik h(x) \sqrt{2-2\sin\theta} e^{ikSx} \equiv
h(x) D_\theta (x),   
\label{42} 
\end{aligned} 
\eeq
and 
\beq C=C_0+C_1   \label{43}\eeq  
where
\begin{align*}  
C_0 &= -\pi L_0^{-1} R_0 D_\theta \\
             C_1 &= -\pi L_0^{-1} R_0 {dI\over dx} .  
\end{align*} 
Therefore to $O(\sigma^2)$ the second moment can be written
\beq  
\begin{aligned}
& m_2(x,y) = \mtil (x,y) +
\psi_0(x) \left<h(y)C_1^*(y)\right> + \left<\psi_1(x) h(y)\right>
C_0^*(y)  \\ 
& +
\psi_0^*(y) \left<h(x)C_1(x)\right> + \left<\psi_1^*(y) h(x)\right> C_0(x)  +
\rho(x-y) C_0(x)C_0^*(y) . \label{44} 
\end{aligned}
\eeq 
Since $\CC=<hC>=<hC_1>$, equation (\ref{44}) can be expressed as
\beq 
\begin{aligned}
 & m_2(x,y) = \mtil (x,y) +
\psi_0(x) \CC^*(y) + \psi_0^*(y) \CC (x)  \\ 
  & ~+ \rho(x-y) C_0(x)C_0^*(y) +
\left<\psi_1(x) h(y)\right> C_0^*(y)
+ \left<\psi_1^*(y) h(x)\right> C_0(x)
. \label{45} 
\end{aligned}
\eeq 
In this equation, only the last two terms remain to be determined.
From (\ref{42}), \hfil\break
$<\psi_1(x) h(y)>$ is just
\beq  \left<\psi_1(x) h(y)\right> = \rho(x-y) D_\theta \label{46} \eeq 
and similarly for $\left<\psi_1^*(x) h(y)\right>$
so that (\ref{45}) becomes
\beq  
\begin{aligned}
& m_2(x,y) = \mtil (x,y) +
\psi_0(x) \CC^*(y) + \psi_0^*(y) \CC (x)  \\
& ~  + \rho(\xi) \Bigl[ C_0(x)C_0^*(y) + D_\theta(x) C_0^*(y)
+ D_\theta^* (y) C_0(x)  \Bigr]  \label{47} 
\end{aligned}
\eeq 
where $\xi=x-y$.

\section{Conclusions}\label{IV}
The parabolic integral equation method has been extended here to allow the
calculation of backscatter of due to a scalar wave impinging on a rough surface
at low grazing angles. The solution is written in terms of a series of
Volterra operators, each of which is easily evaluated, and which
allows examination of multiple scattering resulting from increasing
orders of surface interaction.   Truncation at the first term the
leading forward- and back-scattered components; higher-order multiple
scattering are available from subsequent terms.
The parabolic Green's function is applicable for wave components at low angles
of incidence and scatter, which imply small surface slopes,  but
without restriction on surface heights.
With the additional assumption of small surface heights, analytical
solutions have then been obtained, to second order in height,
for the mean field and its autocorrelation. These provide backscatter
corrections to the solutions given in the purely forward-scattered
case\cite{12,13} with the potential for further insight into the role of
different orders of multiple scattering.  (Small height perturbation
theory derived directly from Helmholtz equation has of course been well
established for many years and yields particularly simple single
scattering results. The results here are from a different perspective;
the first term already includes 'multiple-forward-scattering', and
subsequent terms incorporate back- and forward-scatter contributions
systematically at higher orders.) 
 
In the context of long-range propagation at low grazing angles,
parabolic equation methods remain very widely used. In this regime
the form of the Green's function together with the series
decomposition provide computational efficiency and the means to
extend existing PE methods to include backscatter, 
in addition to yielding tractable analytical results for statistical moments.
These benefits should, nevertheless, be put in context. The computational
advantages of the PE Green's function over the full free space Green's
function are lost in fully 3-dimensional problems (since evaluation of
the 3D PE Green's function is computational expensive),
or those for which wide-angle scatter needs to be taken 
into account. 
 On the other hand there remains a need for further
theoretical understanding of the mechanisms of enhanced and multiple
backscatter, and the approach here may be applied in a more general
setting.    Computational and theoretical results in application to long-range
propagation over rough sea surfaces will appear in a separate paper.

\vfil\break
\appendix
\section*{Appendix}

We can write the expression (\ref{34}) as
\beq  \CC (X,x)= -{ik\over\alpha^3} \cos\theta~ {d\over dx}\int_0^x
g(x,y) H(X,y) dy \label{A.1} \eeq 
where
\beq  g(x,y) = {1\over\sqrt{x-y}}, \label{A.2} \eeq 
\beq  H(X,y)=
\int_y^\infty {e^{ikr}\over\sqrt{y-r}}
\int_0^r {e^{ikSr'}\over\sqrt{r-r'}}
R(X,r') dr'
~ dr ~  , \label{A.3} \eeq 
and $R$ is given by (\ref{35}).
Differentiation with respect to $x$ is carried out
as for the $r$-derivative (equations (\ref{27})-(\ref{33})):
The $x$-derivative is thus expressed as a limit of a finite difference, and the
integral split into three parts,
\beq  \CC (X,x) = -{ik\over\alpha^3} \cos\theta~
\lim_{\epsilon\rightarrow 0} {1\over\epsilon}
 \left[L_1 + L_2 - L_3 \right]  \eeq 
where
\begin{align*} L_1 &=  \int_0^{\epsilon} g(x+\epsilon,y) H(X,y) dy  \\
L_2 &=  \int_\epsilon^{x+\epsilon} g(x+\epsilon,y) H(X,y) dy    \\
L_3 &= \int_0^x g(x,y) H(X,y) dy      \end{align*} 
We thereby obtain
\beq  {d\over dx}\int_0^x
g(x,y) H(X,y) dy = {2\over \sqrt{x}} H(X,y) +
\int_0^x g(x,y) \left[ {dH(X,y)\over dy} - H(X,y) \right] ~dy .\label{A.4} \eeq 
The term $dH/dy$ is then
\beq {dH(X,y)\over dy} = {d\over dy}\int_y^\infty  a(y,r) J(X,r) dr \label{A.5} \eeq 
where
\beq  a(y,r) = {e^{ikr}\over\sqrt{y-r}}, \label{A.6} \eeq 
\beq J(X,r)=
\int_0^r {e^{ikSr'}\over\sqrt{r-r'}} R(X,r') dr' 
\label{A.7} \eeq 
Treating the derivative as before gives
\beq  {dH\over dy} = - \int_y^\infty a(y,r)
\left( {dJ(X,r)\over dr} + ikJ(X,r)\right) ~dr  . \label{A.8} \eeq 
Finally,
\beq  {dJ\over dr}= {d\over dr}
\int_0^r {e^{ikSr'}\over\sqrt{r-r'}} R(X,r') dr'
\label{A.9} \eeq 
from which we similarly get
\beq  {dJ\over dr} =
{R(X,0)\over \sqrt{r}} + \int_0^r {e^{ikSr'}\over\sqrt{r-r'}}
\left\{ ikS R(X,r') + {dR\over dr'}\right\} dr' . \label{A.10} \eeq 
As before (see (\ref{34})) the expression $R(X,0)$ vanishes for large $X$ and
can be dropped.  Successively substituting (\ref{A.8}), (\ref{A.10}),
(\ref{A.3}) and (\ref{A.4}) into (\ref{A.1}),
we eventually obtain
\beq  \left< h(X) C(x) \right> = -{ik\over\alpha^3} \cos\theta~  \times \eeq 
\begin{align} &
\left[ {2\over\sqrt{x}}\int_y^\infty {e^{ikr}\over\sqrt{y-r}} \int_0^r
{e^{ikSr'}\over\sqrt{r-r'}} R(X,r') dr' dr
\right.   \\
& - \left. \int_0^x {1\over\sqrt{x-y}}
\int_y^\infty {e^{ikr}\over\sqrt{y-r}} \int_0^r
{e^{ikSr'}\over\sqrt{r-r'}}
{\cal F}(X,r') ~ dr' ~ dr ~dx \right]
\label{A.11} \end{align} 
where
\beq  {\cal F} = \left\{(1+ik[1+S])R(X,r') + {dR\over dr'}\right\} .
\label{A.12} \eeq 
In this expression, $R$ is given by (\ref{35}), so that
\beq {dR\over dr'} = ikS{d\rho(X-r')\over dr'} - {d^2\rho(X-r')\over dx^2} .
\label{A.13} \eeq 
It is clear then that the correction term introduces a higher-order dependence
on the correlation function.

\vfil\break

\end{document}